\documentclass[aps,twocolumn]{revtex4}
\usepackage{graphicx}
\usepackage{rotating}
\usepackage{dcolumn}
\usepackage{epsfig}

\begin{document}

\title{Global Bifurcation Destroying The Experimental  Torus  $T^2$}
\author{T.   Pereira$^{1,2}$,  M.S. Baptista$^2$, M.B.  Reyes$^1$,
I.L. Caldas$^1$, J.C. Sartorelli$^1$, and J. Kurths$^{2}$}

\affiliation{
$^1$Instituto de F{\'\i}sica, Universidade de S{\~a}o Paulo\\
Caixa Postal  66318, 05315-970 S{\~a}o Paulo, SP, Brasil\\
$^2$Nonlinear Dynamics, Institute of Physics,
University of Potsdam, D-14415, Potsdam, Germany\\
}
\date{\today}

\begin{abstract}
  
  We show experimentally the scenario of a two-frequency torus $T^2$
  breakdown, in which a global bifurcation occurs due to the collision
  of a torus with an unstable periodic orbit, creating a heteroclinic
  saddle connection,   followed by an intermittent behavior.

%PACS:  05.45.Xt, 05.45.-a,  05.45.Pq,02.50.CW
\end{abstract}

\maketitle

Chaotic behavior has been extensively studied in several areas of
physical sciences \cite{Ott}, in economy \cite{economy}, in ecology
\cite{ecology} and in applied engineering \cite{engenharia}.  One
important question toward the understanding of chaotic motion is the
transition from a regular behavior to a chaotic one.  This transition
has been studied in conservative \cite{hamilton} and dissipative
systems \cite{Eckmann,Ott}.

Among the ways chaotic behavior appears are the routes via the
bifurcations of a quasi-periodic attractor with $N$ incommensurate
frequencies, also known as torus $T^N$.  This route for the higher
dimension torus, in which $N \geq 3$, has been well analyzed in Refs.
\cite{Ruelle,Grebogi}. For the low dimension torus $T^2$, chaos might
appear by the Curry-Yorke scenario \cite{cy,Tito,Yu}, and by torus
doubling bifurcations \cite{periodDoubling}. A torus $T^2$ transition
to chaos was also observed in a $CO_2$ laser\cite{labate}

Recently, Baptista and Caldas \cite{Murilo1,Murilo2} have shown the
appearance of chaotic behavior by an abrupt transition directly from
the torus $T^2$, in which there is a collision of the torus with
saddle points \cite{saddle} originating an intermittent behavior. They
argued that there is a mechanism of reinjection associated to a
heteroclinic connection between the central unstable focus and the
saddles, which takes the trajectory back to the vicinity of the focus.
As reported by Letellier et al.\cite{Lethelie}, this phenomenon is not
only restricted to continuous-time systems but it also happens in
spatio-temporal systems.

In this work, we report experimental details of this scenario of $T^2$
breakdown.  We show the global dynamics by identifying the collision
of the torus with the saddle points, and by characterizing the local
dynamics showing the existence of the focus and saddle points, i.e.
the stretching and the folding character of the saddle vicinity, and
the spiraling character of the focus vicinity. Finally, we also
experimentally characterize the observed intermittency.

The forced Chua's circuit is composed of two capacitors, $C_{1}$ and
$C_{2}$, two resistors, $R$ and $r$, one inductor, $L$, and the
non-linear active (piecewise-linear) resistor, $R_{NL}$\cite{Murilo1}, whose
characteristic curve is mathematically represented by:
$i_{NL}(V_{c1})=m_{0}V_{c1}+0.5(m_{1}-m_{0})\{\left|V_{c1}+B_{p}\right|
-\left| V_{c1}-B_{p}\right| \}$, where $V_{c1}$ is the voltage in the
capacitor $C_{1}$ .  The forcing applied to the circuit is of the form
$V_p\sin (2\pi f_p t)$, where $V_p$ is the amplitude and $f_p$ is the
frequency. More details about this circuit can be found in Ref.
\cite{Murilo-Epa}.

The circuit is constructed such that the characteristic curve presents
$m_{0}$=-0.539(3)$mA/V$, $m_{1}$=-0.910(2)$mA/V$, and $
B_{p}$=1.200(4)$V$. The number between brackets indicate the deviation
in the last digit. The components of the circuit are $
C_{1}=0.0047\protect\mu $F, $C_{2}=0.052\protect\mu $F, $R\in [1.0 ,
1.7] K\Omega $, $ L = 9.2$mH, and $r = 10 \Omega$.  We set the
resistor $R$ to obtain the regime of a Double-Scroll-like attractor,
and then we introduce the perturbation destroying this attractor,
which gives place to the attractors observed in this work.  We use a
Tektronix AFG320 function generator connected in parallel to the
resistor $r$, to introduce the forcing in the circuit.  Data
acquisition is performed using an AT-MIO 16E1 National Instruments
board (12 bits) with a sampling of $\delta$=1/250kHz.

We acquire the variable $V_{c_1}(t=i \delta)$, with $i$ representing
the number of acquired data points. Next, we identify all its local
maxima $V_{c_1}(n)$, where the $n$ index represents the number of
maxima.  This time series $V_{c1}(n)$ is used to reconstruct the
attractor $V_{c_1}(n) \times V_{c_1}({n+p})$, using time-delay
embedding coordinates, with $p$ representing the time-delay for the
embedding space. The value of $p$ is chosen conveniently according to
the geometry of the attractor studied, here $p=2$.  By working with
this embedding, we can better visualize the attractors and the
manifolds since this space has one dimension less than the usual phase
space.

\begin{figure}[!h]
  \centerline{\hbox{\psfig{file=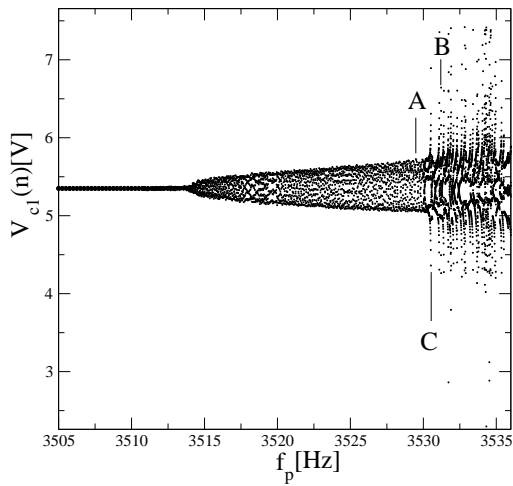,height=8cm}}}
\caption{  Bifurcation   diagram of the  forced Chua's circuit,
  for $V_p=0.7$ V.  One sees a fixed point that bifurcates by a Hopf
  bifurcation into a torus $T^2$. At a critical parameter
  $f_p=3530.6$Hz, indicated by the letter C, a torus-saddle collision.
  The attractors indicated by the letters A and B are shown in Figs.
  \ref{saddle} and \ref{abrupt1} respectively}
\label{diagrama_abrupto}
\end{figure}

We use as control parameters the voltage $V_p$ and frequency $f_p$ of
the forcing.  As we change these parameters several windows of
periodic motion appear.  In particular, we explore the system close to
the periodic window at $V_p = 0.7 V$ and $f_p = 3500$Hz.  For $V_p =
0.7 V$ fixed, we change $f_p$ within the interval $[3505,3537]$ Hz and
build the bifurcation diagram as shown in Fig.
\ref{diagrama_abrupto}.  The periodic window remains up to $f_p = 3514
$ Hz, when a Hopf bifurcation occurs.  Further increasing the control
parameter, at $f_p = 3530.6$Hz (indicated by letter C) a sudden change
in the system dynamics happens, the destruction of the torus by a
torus-saddle collision, generating an intermittent behavior.
Close to the Hopf bifurcation point, the torus has a circular shape.
As the parameter is increased the torus shape becomes like a pentagon.
This change happens when the torus grows and gets close to the five
fixed points corresponding to a period-5 saddle orbit.

The existence of the saddle points, near this distorted torus, can be
verified by introducing a small perturbation in the forced Chua's
circuit, e.g. changing the magnetic flow in the inductor.  This
perturbation sets the trajectory apart from the
attractor\cite{robustness}.

\begin{figure}[!h]
  \centerline{\hbox{\psfig{file=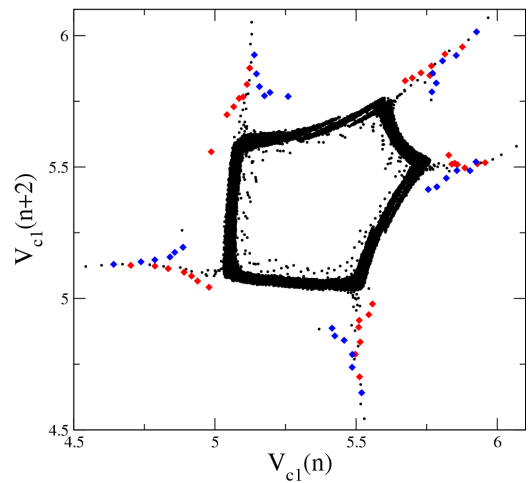,height=8cm}}}
\caption{ \label{saddle} (Color online) The pentagon-shape structure is the torus $T_2$,
  and with the small filled circles we show orbits that are placed in
  the vicinity of the saddle as we perturb the circuit.  This takes
  the trajectories away from the torus and reveals the structure of
  the phase space near the torus, i.e., the existence of the saddles.}
\end{figure}

\begin{figure}[!h]
  \centerline{\hbox{\psfig{file=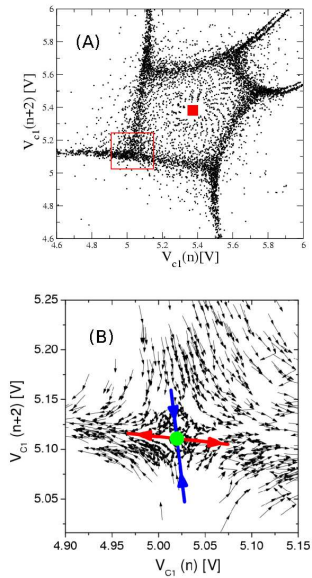,height=8.0cm}}}
\caption{ \label{abrupt1} (Color online) Fig \ref{abrupt1}(A) The chaotic 
  attractor born from a global bifurcation, the torus-saddle
  collision, for $f_p = 3531$Hz.  The square in the center of the
  broken torus represents the unstable focus point.  Fig.
  \ref{abrupt1}(B), we analyze the dynamics inside the small square in
  Fig \ref{abrupt1}(A).  The arrows connect a given point with its 5th
  iterate. One can see the saddle structure represented by the arrows,
  the unstable (stable) manifold is revealed by looking at the
  direction of the vectors. The filled circle indicates the saddle
  point position and the oversized arrows indicate the unstable and
  stable manifolds of the saddle point.}
\end{figure}

With this procedure we are able to place the trajectory in the
vicinity of the saddle points, as well as, in different parts of the
phase space.  We repeat this perturbation until we have a picture of
the phase space in which we can clearly see the saddle points. Those
are identified by selecting trajectory points that after
$p$-iterations are mapped close by.  In Fig. \ref{saddle}, we show the
torus that resembles a pentagon and the perturbed trajectories that
are placed in the vicinity of the saddles. After reaching the saddle
vicinities, the trajectory goes either toward the torus or away from
it, by the stable or the unstable manifolds of the saddle points
(filled circles), respectively.
At a critical parameter $f_p=3530.6$ Hz, the torus collides with the
external saddle points, it breaks and an intermittent behavior takes
place.  Analyzing the destroyed torus, depicted in Fig.
\ref{abrupt1}(A), we can still detect the presence of the saddle points,
which indicates that at the collision of the saddle with the torus
$T^2$ a heteroclinic saddle connection is formed. That is done by
analyzing the local geometry of the mapping around the saddle points,
i.e. observing the return of a set of initial conditions around these
saddles, after 5 iterations.  In Fig. \ref{abrupt1}(B), inside the
square shown in the Fig. \ref{abrupt1}(A) we connect with arrows the
points in the set of initial conditions and the corresponding
returning points. We selected the shortest arrows in order not to
overwhelm the picture, revealing the saddle.

We also see that the dynamics is still related to the torus. In fact,
the trajectories behave as if they were in a quasi-periodic
oscillation, near the saddle points, for a while.  Then, they go away
from the saddle vicinity along the saddle unstable manifolds, spending
some time far away from the region delimited by the saddles.  Finally,
they are eventually reinjected in the unstable focus vicinity,
spiraling toward the saddle points.

The unstable focus, in the attractor shown in Fig. \ref{abrupt1}(A), was
obtained by the Fixed Point Transform \cite{pso}, which consists in
transforming the original data set, such that the transformed data is
concentrated on the periodic orbits. The histograms of the transformed
data presents sharp peaks at the location of the unstable periodic
orbits, which allow us to estimate their positions.
The trajectories in the branches of the broken torus, the extended
structures in Fig. \ref{abrupt1}(A), are governed by the unstable
manifold of the saddle points.  To see this, we evolve in time a set
of points in the branch, close to the saddle, and after $k \times 5$
iterations, with $k$ small, this set returns to the branch, but
stretched.  Following the consecutive iterates of the branches, we
notice that the points on the branches are eventually mapped in the
vicinity of the focus.  The scaping of the trajectory from the saddles
(chaotic bursts) and the subsequent reinjections around the focus
generate the intermittent behavior.

\begin{figure}[!h]
  \centerline{\hbox{\psfig{file=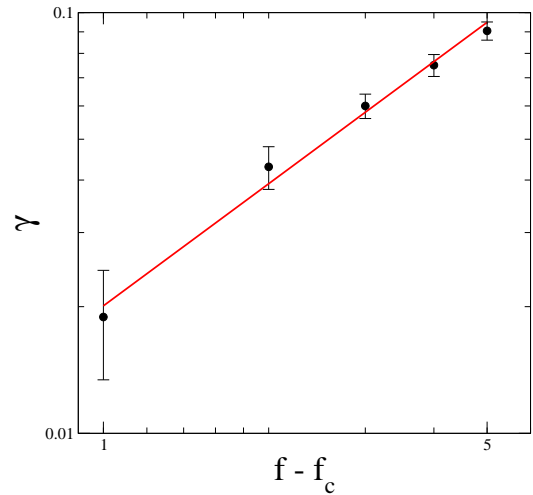,height=8cm}}}
\caption{\label{slaw} We show $\gamma \times |f - f_c|$, in a $log \times log$ 
  graph.  By fitting the data set, we have that the average laminar time
  scales as $\tau \propto |f - f_c|^{\mu}$, where $\mu = 0.96$.}
\end{figure}

In experimental situations, measuring the average laminar time, namely
$\tau$, can be rather problematic, as noted in Ref. \cite{Frank},
since one might mislead the determination of the laminar period, by
working with variables that are in a projection of the full space. It
is better to compute a more robust quantity called $\gamma$. This
quantity measures the fraction of the time of the chaotic burst
$\gamma=t_{burst}/t_{total}$ and, as the chaotic burst is very
pronounced, the error in the statistics is smaller.  Assuming $\tau
\propto \epsilon^{-\mu}$, the variable $\gamma$ is related to $\tau$
by $\gamma = \left( 1 + \frac{d}{\epsilon^{\mu}}\right)^{-1}$ with $d$
being a constant (see Ref. \cite{Frank} for more details). Since $ 1
<< d / \epsilon^{\mu}$, we can write $\gamma \propto \epsilon^{\mu}$.
From our experimental results we have that $\mu = 0.96$ (Fig.
\ref{slaw}).
Next, we estimate the Lyapunov exponents of our data through the
Eckmann-Ruelle technique \cite{Eckmann}, by using the two-dimensional
embedding $\{ V_{c1}(n), V_{c1}(n+2)\}$. The spectra of the Lyapunov
exponents, for the forced Chua's circuit has the form $\Lambda
=(\lambda_1,0,\lambda_2, - c)$.  The null exponent corresponds to the
direction of the flow, and the $-c$ exponent corresponds to the strong
stable foliation.  The other two exponents, $\lambda_1$ and
$\lambda_2$, correspond to the stretching and the folding dynamics of
the return map $V_{c1}(n) \times V_{c1}(n+2)$.  The spectrum that
corresponds to the attractor of Fig.  \ref{abrupt1} is $\lambda_1 =
0.30 \pm 0.08$ and $\lambda_2 = -0.10 \pm 0.04$, with one positive
Lyapunov exponent, and therefore chaotic behavior.
\begin{figure}[!h]
  \centerline{\hbox{\psfig{file=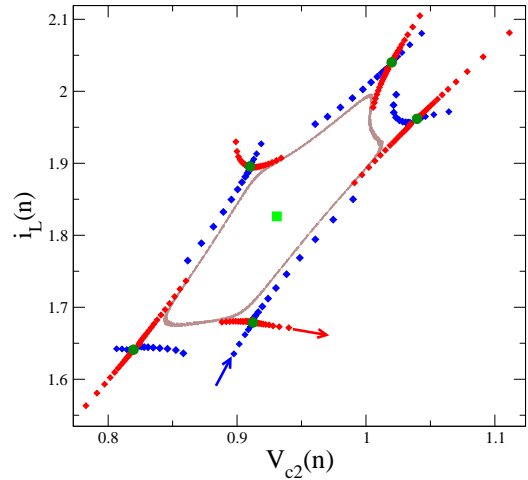,height=8cm}}}
\caption{\label{manifolds} (Color online) We show the torus (gray dots), the unstable focus 
  (light gray square), the saddle points (filled circles), and a
  picture of the manifolds (diamonds) of the saddle points. These
  manifolds are depicted by iterating a set of initial conditions very
  close to one saddle point. The stable manifold of the saddles are
  indicated by dark diamonds (see the dark arrow), and the unstable
  manifolds are indicated by the gray diamonds (see the gray arrow).}
\end{figure}

In order to have a better picture of the way the saddle points collide
with the torus we perform numerical analysis of the circuit equations,
with parameters that adequately reproduce the experimental scenario.
The circuit equations can be obtained by applying the Kirchoff's laws.
The resulting state equations are given by : $ C_{1}\frac{dV_{c1}}{dt}
= \frac{1}{R}(V_{c2}-V_{c1})-i_{NR}(V_{C1}), \,
C_{2}\frac{dV_{c2}}{dt} = \frac{1}{R} (V_{c1} -V_{c2})+i_{L}, \,
L\frac{di_{L}}{dt}= -V_{c2} -V_p \sin{\phi}, \, $ and
$\frac{d\phi}{dt} = 2 \pi f_p $, where $V_{c1}$ and $V_{c2}$ are the
voltage in the capacitors $C_{1}$ and $C_{2}$, respectively, and
$i_{L}$ is the electric current across the inductor $L$.

The parameters used in the numerical simulation are $C_{1}=0.1$,
$C_{2}=1.0$, $L=1/6$, $\frac{1}{R}=0.575$, $m_{0}=-0.5$, $m_{1}=-0.8$,
and $ B_{p}=1.0$.  The experimental scenario can be reproduced if we
set the perturbation amplitude at $V = 0.234$ and change the frequency
in the interval $f_p =[0.1775,0.1795]$. The numerical analysis are
done in a Poincar\'e section at $V_{c_1} = -1.5$. Plotting $V_{c_2}(n)
\times f_p$ we obtain similar pictures as in the experimental
scenario.

In order to detect the saddle points as well as the focus, we
introduce a numerical technique, which can be adapted to experimental
data series.  The method consists in integrating the equations for
several trials and changing the initial condition in each trial.  The
initial conditions are randomly chosen within a volume centered at a
point $p_0$ in the chosen Poincar\'e section of the phase space. The
point $p_0$ has to be close to the searched period-$n$ unstable orbit,
the distance was set to be smaller than $10^{-8}$.  Applying this
method we were able to detect the unstable periodic orbits of the
flow, which gives place to the saddle points in the Poincar\'e
section. The unstable orbit exists for all parameters within $f_p
=[0.1775,0.1795]$ and $V = 0.234$.

At a particular parameter $f_p =0.17893$, we manage to get a local 
illustration of the manifolds of the saddle points. For that, we
evolve forward and backward a
small set of initial condition very close to a saddle point, 
just a few returns to the section. Then,
having a better picture of the local structure of the manifolds, we
choose new initial conditions, which are close to the manifolds and
iterate them.  The result is shown in Fig.  \ref{manifolds}, with
the saddles indicated by the filled circles, coexisting with the
pentagon-shape torus $T^2$ in gray dots.  In this figure one can
clearly see the distortion of the torus $T^2$ due the manifolds of the
saddle points.

Concluding, we analyze experimentally the scenario of a two-frequency
torus breakdown by a global bifurcation.  We show: $(i)$ The existence
of the saddle points; $(ii)$ The existence of the focus point; $(iii)$
a local picture of the manifolds of the saddle points; $(iv)$ The
verification of a power scaling law for the average laminar length,
$\tau \propto |f - f_c |^{-0.96}$.

\textbf{Acknowledgment}: We thank M. J. Sotomayor and A.
J. P. Fink for useful discussions. We acknowledge support from the
the Alexander Humboldt Foundation (MSB), 
``Helmholtz Center for Mind and Brain Dynamics'' (TP and JK), and 
FAPESP.


\begin{thebibliography}{999}
\bibitem{robustness} The perturbation is introduced in the oscillator
  set with a parameter far enough from the bifurcation point in the
  parameter space. In this region the phase space is robust under
  small perturbations.
\end{thebibliography}
\end{document}